\documentclass[a4paper]{jpconf}
\usepackage{graphicx}
\begin{document}
\newcommand{\zncu}{ZnCu$_3$(OH)$_6$Cl$_2$ }

\title{Thermodynamic Properties of Kagome Antiferromagnets with
different Perturbations}

\author{Rajiv R. P. Singh$^1$ and Marcos Rigol$^2$}

\address{$^1$ University of California, Davis, CA 95616}

\address{$^2$ Georgetown University, 37th and O St NW, Washington, DC 20057}

\ead{singh@physics.ucdavis.edu}

\begin{abstract}
We discuss the results of several small perturbations to 
the thermodynamic properties of Kagome
Lattice Heisenberg Model (KLHM) at high and intermediate temperatures, 
including Curie impurities, 
dilution, in-plane and out of plane Dzyaloshinsky-Moria (DM) anisotropies and exchange
anisotropy. We examine the combined role of Curie impurities and
dilution in the behavior of uniform susceptibility. We also
study the changes in
specific heat and entropy with various anisotropies.
Their relevance to newly discovered materials
\zncu is explored. We find that the magnetic susceptibility is well
described by about $6$ percent impurity and dilution. We also find that
the entropy difference between the material and KLHM is well described
by the DM parameter $D_z/J\approx 0.1$.
\end{abstract}

The discovery of Herbertsmithite compounds \zncu has
led to considerable excitement \cite{shores}. It is an example of the long-sought
antiferromagnetic spin-half Kagome Lattice Heisenberg Model (KLHM). 
Ideally speaking, the layered material
consists of spin-half copper atoms arranged in structurally perfect
Kagome planes, which are separated by non-magnetic planes. The latter contain
only zinc atoms as the transition metal. This leads to the possibility
of observing ideal KLHM behavior in the material. Indeed, in
several ways the experimental
observations \cite{helton,ofer,mendels,imai,shlee,bert,zorko,olariu} 
appear promising for one of the highly anticipated but
so far elusive states of matter, a two-dimensional spin-liquid with
deconfined gapless spin-half excitations.

However, the situation is unsettled on both the experimental and
theoretical fronts. On the theoretical front, it remains unclear
that the ground state of KLHM is a gapless spin-liquid
with deconfined spin-half excitations \cite{ran,hermele,hermele2}. Evidence for absence
of long-range order and a gap in the spin excitation spectrum
has come from several numerical studies \cite{zeng,singh-huse1,leung,lecheminant,waldtmann,sindzingre,mila,singh-huse2,singh-huse3,jiang}. Furthermore, a
Valence Bond Crystal phase has been proposed \cite{marston,nikolic}, which appears to have a lower
energy \cite{singh-huse2,singh-huse3} than the proposed variational wave functions
for the spin-liquid phases \cite{ran}. So the question remains, is
the spin-liquid phase, either with Fermi points or with a true
Fermi surface for spinons realized for KLHM? And, can it be
stabilized by adding other smaller interactions to the KLHM \cite{ma}?
This remains an active area of theoretical investigations.

On the experimental front, a key question is, do the experimental observations
reflect the behavior of an ideal KLHM, or are they dominated
by various perturbations, including quenched impurities \cite{shores,helton,ofer,mendels,imai,shlee,bert,zorko,olariu}. The exchange
constant of the material is approximately $200 K$ \cite{rigol1,misguich}, yet the magnetic susceptibility
continues to increase below 1K and saturates at even lower temperatures to
a very large value, in dimensionless terms almost two orders of magnitude larger
than the known susceptibility for the square-lattice Heisenberg Model. Furthermore,
the specific heat appears linear or even sub-linear at low temperatures,
and is strongly magnetic-field dependent \cite{helton}.
However, one should keep in mind that this very low temperature behavior 
reflects a tiny fraction of
the full spin entropy of $\ln{(2)}$ \cite{misguich}. 
Hence, its relevance to the KLHM is doubtful.
Among other experimental results, there is no signature of a spin-gap
either in Neutron scattering, or in NMR or $\mu$SR measurements \cite{helton,ofer,mendels,imai,shlee,bert,zorko,olariu}. Before
interpreting the experiments in terms of various proposed ground state
scenarios for the KLHM, it is important to ascertain how much of the
behavior is intrinsic to the Kagome system and how much of it arises from
quenched impurities.

There is growing evidence that these 
materials have significant antisite disorder \cite{imai,shlee,bert,zorko,olariu}.
This is related to interchange of copper and zinc atoms between the kagome
and non-magnetic planes. It has been estimated that upto 6-10 percent of the
copper atoms may be replaced by zinc atoms and themselves end up in the zinc planes.
This can cause a large density of nearly free spins, which would be only
weakly coupled to the spins in the kagome planes. Furthermore, it leads
to dilution in the Kagome planes, a quenched disorder, which can significantly
affect the thermodynamic properties in the plane, especially if large
unit cell Valence Bond Crystal phases are relevant to the pure material.

The Kagome Lattice lacks inversion symmetry through midpoints
of bonds and hence, always allows for
Dzyaloshinsky-Moria (DM) interactions \cite{dzy,moriya}, involving cross products
between spins. In a structurally perfect purely two-dimensional
Kagome-Lattice, the DM interactions are represented by a vector $D_z$, which
points out of the Kagome plane. However, in the real material, the three
dimensional embedded structure has sufficiently low symmetry to allow
two independent DM parameters. A $D_z$, which points out of the plane
and a $D_p$, which points in the plane, is perpendicular to the bonds
and hence rotates from bond to bond.
A priori, either of these DM terms could be dominant in the materials. 
Recent ESR experiments have been interpreted in terms of a dominant $D_z$
term, which is about ten percent of the exchange interaction $J$ \cite{zorko}.

The low temperature behavior of the KLHM is characterized by high
near degeneracy between many putative ground states. Thus the
system is clearly going to be highly sensitive to small perturbations.
The $D_z$ interactions lift the high ground state degeneracy at
the classical level, leading to planar order with definite 
chirality \cite{elhajal}.
Recently, these $D_z$ interactions have been studied numerically 
for the spin-half system by Cepas et al \cite{cepas},
who find a phase transition from a non-magnetic
phase to a magnetically ordered phase at a critical ratio of $D_z/J$ of
approximately ten percent. This puts the material \zncu close to
this quantum critical point.

In contrast to the $D_z$ interactions, the $D_p$ interactions are much more
complex. At the classical level, they can favor a weakly ferromagnetic
canted state \cite{elhajal}. For the quantum system, it is likely that they may not lead
to long-range order even for sufficiently strong $D_p/J$ ratio. In fact,
the finite temperature behavior of the entropy function shows \cite{rigol2} that while 
an increasing $D_z$ leads to a clear reduction of entropy with respect to
the pure KLHM, an increasing $D_p$ even to values as large as $J/4$ hardly
changes the entropy function. Numerical studies also found that $D_p$
interactions can lead to enhanced ferromagnetic susceptibilities, which
could play a role in explaining the experimental observations \cite{rigol1,rigol2}.

In addition to DM anisotropy, the possibility of relatively strong Ising
anisotropy in these systems has also been suggested \cite{chern,ofer2}. The pure Ising model
on the Kagome Lattice is exactly soluble and leads to a large ground state
entropy. Since a fraction of the ground states have a finite magnetization,
this system has a divergent Curie-like susceptibility as $T\to 0$. Moving
away from the Ising model, by adding $XY$ exchange terms leads
to (i) a lifting of the ground state degeneracy, (ii) a robust second peak in the
specific heat at low temperatures, and (iii) a cut-off for the Curie susceptibility
as $T\to 0$ \cite{rigol3,rigol4}. Such a behavior has indeed been reported in the experiments.
Besides the high temperature susceptibility of oriented samples shows vastly
different Curie-Weiss temperatures along and perpendicular to the kagome planes \cite{ofer2}.
Thus, substantial Ising anisotropy cannot be ruled out for these materials.

In this paper, our focus is going to be on the thermodynamic properties
at relatively high temperatures, which can be studied in
a controlled manner by High Temperature Expansions (HTE) \cite{elstner},
Numerical Linked-cluster Expansions (NLC) \cite{rigol3,rigol4}, as well
as by Exact Diagonalization (ED) of finite systems \cite{misguich}.
In an earlier paper, we had studied various perturbations individually
including DM and exchange anisotropy, impurities and dilution \cite{rigol2}.
There, our assumptions were
that much of the observed behavior was intrinsic to the
materials and was not due to quenched impurities and that anisotropies
were relatively small, that is, the system was close to being KLHM. 
Thus, while we
did look at impurity, dilution and small Ising and XY anisotropy, 
we focused primarily on the
DM interactions as the primary reason for the large increase
in the magnetic susceptibility. This led to suggestions of 
large DM interactions, which were primarily of $D_p$ character.

\begin{figure}
\begin{center}
\includegraphics[width=24pc]{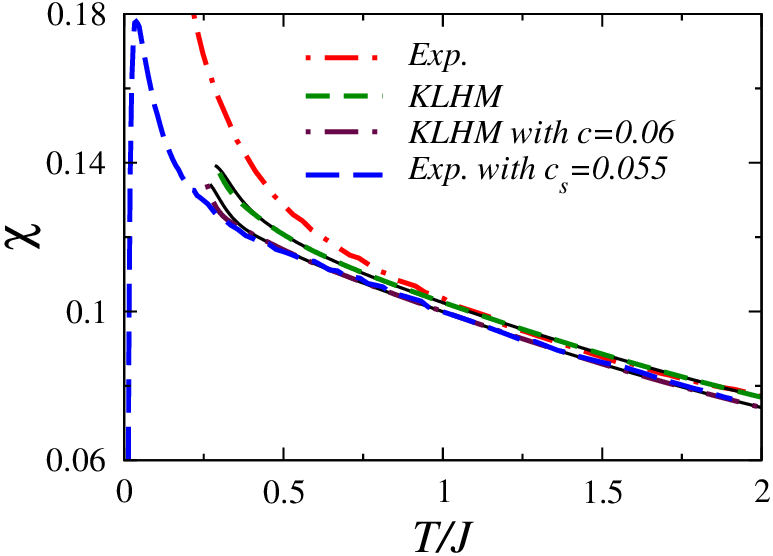}
\end{center}
\caption{\label{Impurity} Experimental susceptibility (Exp.) of the material \zncu compared
with those of the Kagome Lattice Heisenberg Model (KLHM) with J=170K. Also shown is the
experimental susceptibility after subtraction of Curie terms. The latter are
fit remarkably well with the susceptibility of a diluted KLHM with J=210K.}
\end{figure}

Here we look again at these perturbations,
relaxing the assumption that quenched impurities
are not playing a role. In fact, we believe, quenched impurities do play an
increasingly important role at low temperatures. Unfortunately, this makes it difficult
to select a unique set of parameters from a comparison between theory and
thermodynamic experimental properties.
We can combine free spins
with dilution to get an excellent agreement with the susceptibility data
at intermediate and high temperatures, without the need for any anisotropies.
This is shown in Fig.~1, where one can see that when about
six percent free Curie impurities are subtracted from the
experimental data, it agrees remarkably well with the susceptibility of
the KLHM with about six percent dilution \cite{rigol2}.
It is not useful to add further DM
interactions to study the susceptibility with impurity, 
dilution and DM interactions as proliferation of free parameters
makes such an exercise meaningless.
However, we should note that impurity plus dilution
cannot be the full answer to \zncu down to lowest temperatures
as ultimately the Curie behavior goes away and the susceptibility
saturates. The absence of Curie-like growth shows up in the subtracted
curves in Fig.~1 as a rapid downturn in the susceptibility. This quenching of the
impurity susceptibility requires a coupling between the impurity spins and
the rest of the system and cannot be understood if the rest of the system has a spin-gap.

The impurity spins should be nearly free at temperatures above their coupling scale. In this
case, the temperature dependence of the
specific heat and entropy should come from the rest of the system.
We will try to use this to get a handle on the anisotropies.
There has been a long theoretical debate about the specific heat of the pure KLHM
and whether it has a second low temperature peak in addition to the high temperature
peak associated with short-range order \cite{elstner,elser}. This issue is not settled yet. For KLHM
the extrapolation of High Temperature Expansion (HTE) by Misguich and Sindzingre
allows one to reliably calculate entropy down to $T/J=0.06$ \cite{misguich}, which is considerably lower than
what we can get reliably by Numerical Linked Cluster (NLC) expansions. This is because the
extrapolation method developed by Bernu and Misguich \cite{misguich2} allows them to incorporate
a lot of information about the $T=0$ properties of the system into the HTE study.
Hence, in this case, we simply quote several results from their paper. 
First, around $T/J=0.1$ the specific heat 
of the KLHM, per copper site is $0.1 $ (We take $k_B=1$). Second, integrating the experimental specific
heat from $T=0$ to obtain the entropy as a function of temperature shows that at $T/J=0.06$
the entropy of the experimental system falls below that of KLHM by at least $0.05 $.
Misguich and Sindzingre suggest that this implies that additional terms in the experimental
material help quench part of the large low temperature entropy of the system.
The issue of the lower temperature peak in specific heat of KLHM
is sensitive to assumptions about the low temperature
behavior of the model and is not resolved in their study.

\begin{figure}
\begin{center}
\includegraphics[width=24pc]{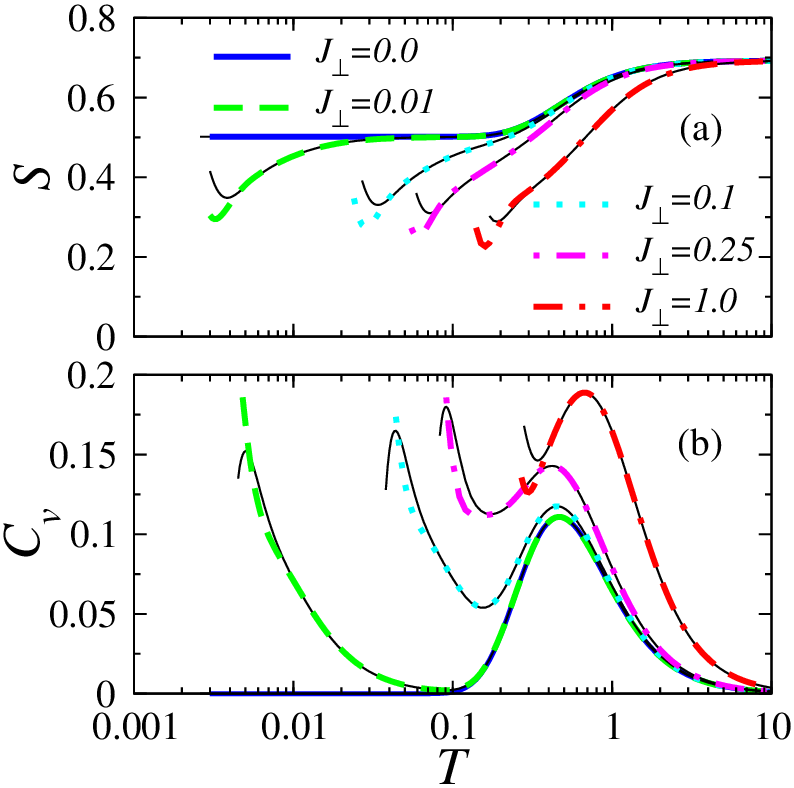}
\end{center}
\caption{\label{HI-model} Entropy and Specific Heat, per spin, of the Heisenberg-Ising Model on
the Kagome Lattice for several values of $J_\perp$ with $J_z=1$.} 
\end{figure}

In Fig.~2, we show the entropy and specific heat of the Heisenberg-Ising model
on the kagome lattice as obtained by NLC \cite{rigol3,rigol4}. While for the Heisenberg model, the existence of a 
two-peaked specific heat remains under debate, there is no question that such a behavior
arises for the Heisenberg-Ising model. For the pure Ising model, the entropy saturates
to its ground state value at relatively high temperatures and the specific heat becomes
exponentially small at lower temperatures. If we add a small XY coupling ($J_\perp$) to
the Ising model, at high temperatures the behavior resembles that of the Ising model.
But at lower temperatures, when $J_\perp$ becomes relevant, the specific heat develops
a second peak and the ground state entropy is lifted by the perturbation. For a range of
large anisotropy, the specific heat has a minimum around $T=J/10$. 

\begin{figure}
\begin{center}
\includegraphics[width=24pc]{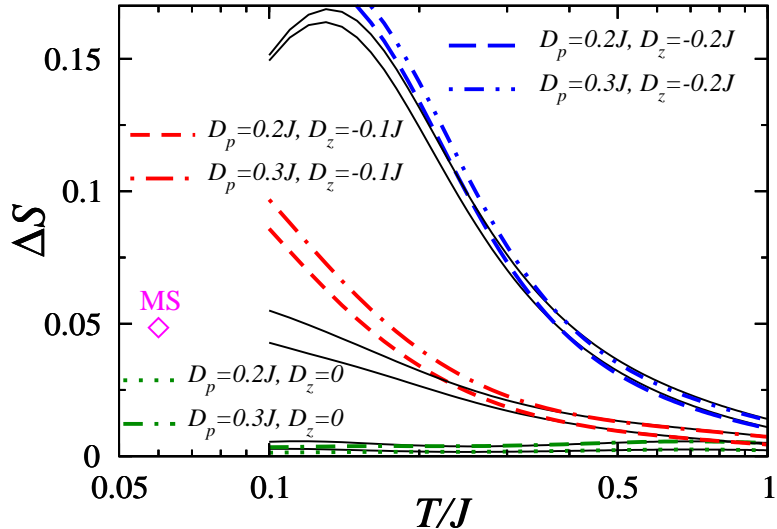}
\end{center}
\caption{\label{Entropy} Entropy reduction with respect to KLHM
due to DM interactions, for several values of the DM parameters. 
Two curves are shown for each parameter set.
The thick dashed lines are from exact diagonalization of a 15-site
cluster, whereas the thin black lines are from exact diagonalization
of a 12-site cluster.
The diamond represents the entropy reduction in \zncu inferred
from the study of Misguich and Sindzingre.}
\end{figure}

From an experimental point of view it has been very difficult to obtain the specific heat
at temperatures above $10K$ (roughly $J/20$). This is because, the phonon contributions
become very large above this temperature and it is difficult to reliably subtract them.
However, such a subtraction has been attempted by Helton and Lee \cite{helton2} and their data suggests
that the specific heat has a minimum around $J/10$. However, the experimental behavior
cannot be taken as evidence for Ising anisotropy for two reasons. First, as seen in
Fig.~2, Ising anisotropy increases the entropy at any temperature with respect to
the Heisenberg model, where as the experimental entropy is lower. And, second, below
the peak the specific heat of the Heisenberg-Ising model rises to a level above $0.1$,
whereas the experimental value is only about $0.03$ \cite{misguich,helton2}.

To look for reduction in entropy with respect to KLHM, we turn to the
DM anisotropies. We consider a whole range of DM parameters ($D_p$ and $D_z$).
For each case, we use exact diagonalization of small clusters of size 12 and 15
to obtain the entropy function \cite{rigol2}. For any temperature, we define the reduction in
entropy with respect to the Heisenberg model as
\begin{equation}
\Delta S=S(D_z=0,D_p=0)-S(D_z,D_p).
\end{equation}
Note that $\Delta S>0$ means the entropy of the system with the DM interactions
is lower than that of KLHM. The quantity $\Delta S$ is plotted in Fig.~3. The
experimental value inferred in the study of Misguich and Sindzingre \cite{misguich} is shown
by a diamond. The main message from this plot is that the reduction in entropy
is very insensitive to $D_p$ and depends primarily on $D_z$. So, if we assume that
the reduction in entropy is due to DM interactions, a $D_z$ value of order ten percent
is likely. This is consistent with the experimental finding.

In conclusions, in this paper we have looked at the role of various perturbations to KLHM
in the finite temperature thermodynamic properties and their relevance to \zncu. The
susceptibility data at intermediate and high temperatures can be rather well fit with
six percent Curie impurities along with a six percent dilution in the Kagome planes.
The reduction in entropy with respect to KLHM below $T/J=0.1$ suggests $D_z/J\approx 0.1$ in these
materials. These findings are consistent with other experimental studies. The very low
temperature behavior of these materials are likely dominated by impurities, which
are weakly coupled to the system. Together with the anisotropies considred here,
they may lead to a very different ground state than the one expected for the pure KLHM.
This is beyond the scope of the current work.

We thank A. Keren, J. Helton and Y. Lee for useful discussions and for sharing their 
data with us.

\end{document}